\documentstyle[multicol,prl,aps,epsf]{revtex}

\begin{document}

\title{Breakdown of the resistor model of
CPP-GMR in magnetic multilayered nanostructures}
\author{S. Sanvito\thanks{e-mail: sanvito@dera.gov.uk},}
\address{School of Physics and Chemistry, Lancaster University,
Lancaster, LA1 4YB, UK and \\
DERA, Electronics Sector,
Malvern, Worcs. WR14 3PS UK}
\author{C.J. Lambert\thanks{e-mail:c.lambert@lancaster.ac.uk},}
\address{School of Physics and Chemistry, Lancaster University,
Lancaster, LA1 4YB, UK}
\author{J.H. Jefferson}
\address{DERA, Electronics Sector, Malvern,
Worcs. WR14 3PS, UK}
\date{\today}
\maketitle

\begin{abstract}

We study the effect on CPP GMR of changing
the order of the layers 
in a multilayer. 
Using a tight-binding simple cubic two band model ($s$-$d$),
magneto-transport properties are calculated
in the zero-temperature, zero-bias limit, within the 
Landauer-B\"uttiker formalism. 
We demonstrate that for layers of different thicknesses formed
from a single magnetic metal and multilayers formed from
two magnetic metals, the GMR ratio and its dependence on disorder
is sensitive to the order of the layers.
This effect disappears in the limit of large disorder, where the 
results of the widely-used Boltzmann approach to transport are restored.

\end{abstract}

\begin{multicols}{2}

\vspace{0.3in}


Giant magnetoresistance (GMR)
in transition metal magnetic multilayers \cite{gmr1,gmr2} is a
spin filtering effect which arises when the 
magnetizations of adjacent layers switch from an anti-parallel (AP)
to a parallel (P) alignment. The resistance in the anti-aligned state
is typically higher than the resistance with parallel alignment, the
difference being as large as 100\%. This sensitive
coupling between magnetism and transport allows the development of magnetic
field sensors with sensitivity far beyond that of
conventional anisotropic magnetoresistance (AMR) devices.
In the most common experimental setup, the current flows in the 
plane of the layers (CIP), and the resistance is measured with
conventional multi-probe techniques. Measurements in which the current
flows perpendicular to the planes (CPP) are more delicate because of the
small resistances involved. Despite these difficulties
the use of superconducting contacts \cite{msu}, sophisticated
lithographic techniques \cite{phil}, and electrodeposition
\cite{el1,el2,el3}, makes such measurements possible 
(for recent reviews see references \cite{bau,ans}).

A widely adopted theoretical approach
to GMR is based on the semi-classical
Boltzmann equation within the relaxation time approximation. This
model has been developed by Valert and Fert
\cite{fer1,fert2}, and has the great advantage
that the same formalism describes both CIP and CPP experiments. 
In the limit that the spin diffusion length $l_{\mathrm sf}$ is much larger 
than the layer thicknesses
(ie in the infinite spin diffusion length limit), this model
reduces to a classical two current resistor network, with
additional possibly spin-dependent scattering
at the interfaces \cite{msu2}.
Despite the undoubted success of this description
recent experiments \cite{chris2,msuord} have drawn attention to the
possibility of new features which lie outside
the theory.
Two important and central predictions of this model are that the
CPP GMR ratio is independent of the number of bilayers in the case that the 
total multilayer length is not constrained to be constant, and 
furthermore is independent of
the order of the magnetic layers in the case of different 
magnetic species.
An apparent violation of the first prediction has been observed in 
CIP and CPP measurements \cite{chris,chris2},
and of the second prediction in CPP measurements
\cite{chris2,msuord}. However a convincing
theoretical explanation is lacking.

\begin{figure}
\narrowtext
\epsfysize=5cm
\epsfxsize=6cm
\centerline{\epsffile{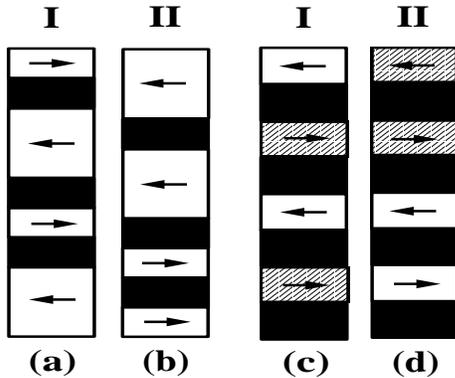}}
\caption{AP configuration for type I and type II multilayers described in the text.
Figures (a) and (b) refer to the cases of thick and thin Co layers with different
coercive fields. Figures (c) and (d) refer to different magnetic metals
coupled through non-magnetic spacers via exchange coupling. The black
blocks represent Cu, the white Co and the hatched Fe$_{72}$V$_{28}$. The arrows indicate
the direction of the magnetizations. Note the difference in the case of
figure (b) where the AP alignment occurs between the two halves of
the multilayer.
}
\label{ord-sc}
\end{figure}

The aim of this letter is to provide a quantitative description of
the breakdown of the resistor
model in diffusive CPP multilayers in the limit of infinite spin-relaxation
length. To illustrate this breakdown,
consider a multilayer consisting of two independent building blocks, namely
a (N/M) and a (N/M$^\prime$) bilayer, where M and M$^\prime$ represent 
magnetic layers of different materials or of the same material but with
different thicknesses
and N represents normal metal `spacer' layers.
From an experimental point of view M and M$^\prime$ must possess different 
coercive fields, in order to allow AP alignment. In the case of
\cite{msuord} this is achieved by considering respectively Co and 
Ni$_{84}$Fe$_{16}$ layers with
Ag as non-magnetic spacer, while in \cite{chris2} both the layers 
are Co (with Cu as spacer) 
but with different thicknesses (respectively 1nm and 6nm).
Two kinds of multilayer can be deposited. The first, that we 
call type I, consists of a (N/M/N/M$^\prime$)$\times \mu$ sequence 
where the species M and M$^\prime$ are separated by an N layer
and the group of four layers is repeated $\mu$ times.
The second, that we call type II, consists of a
(N/M)$\times \mu$(N/M$^\prime$)$\times \mu$ sequence, where the 
multilayers (N/M)$\times \mu$ and (N/M$^\prime$)$\times \mu$
are arranged in series.
If the coercive fields of M ($H_{M}$) and M$^\prime$ ($H_{M^\prime}$) 
are different (eg $H_{M}<H_{M^\prime}$) and if N is
long enough to decouple adjacent magnetic layers, the AP
configuration can be achieved in both type I and type II multilayers
by applying a magnetic field $H$ whose intensity is
$H_{M}<H<H_{M^\prime}$. The AP configuration
is topologically different in the two cases, because 
in type I multilayers
it consists of AP alignment of adjacent magnetic layers
(conventional AP alignment), while in
type II multilayers it consists of the AP alignment between the 
(N/M)$\times \mu$ and (N/M$^\prime$)$\times \mu$ portions of the multilayer,
within which the alignment is parallel (see figure \ref{ord-sc}a
and figure \ref{ord-sc}b).
From the point of view of a resistor network description
of transport, the two configurations are equivalent, because they
possess the same number of magnetic and non-magnetic layers, and
the same number of N/M and N/M$^\prime$ interfaces. Hence the GMR ratio
must be the same.
In contrast the GMR ratio of type I multilayers is found experimentally 
to be larger 
than that of type II multilayers \cite{chris2,msuord}, and the difference
between the two GMR ratios increases with the number of bilayers.
In the case of \cite{chris2}
the GMR ratio of both type I and type II multilayers increases with the number 
of bilayers, which again lies outside the resistor
network model.

In this Letter we demonstrate for the first time
that a description which incorporates
phase-coherent transport over long length scales
can account for such experiments. 
To illustrate this we have simulated type I 
and type II multilayers using a Co/Cu system with
different thicknesses for the Co layers, namely
$t_{\mathrm Cu}=10$AP, $t_{\mathrm Co}=10$AP, $t_{\mathrm Co}^\prime=40$AP.
The technique for computing transport properties
is based on a three dimensional simple cubic
tight-binding model with nearest neighbor couplings 
and two degrees of freedom per atomic site. 
The general spin-dependent Hamiltonian is

\begin{equation}
H^\sigma=\sum_{i,\alpha}\epsilon^{\alpha\sigma}_{i}
c_{\alpha i}^{\sigma\dagger} c^\sigma_{\alpha i}
+\sum_{i,j,\alpha\beta}\gamma^{\alpha\beta\sigma}_{ij}
c_{\beta j}^{\sigma\dagger} c^\sigma_{\alpha i}
\;{,}
\label{spham}
\end{equation}

where $\alpha$ and $\beta$ label the two orbitals
(which for convenience we call $s$ and $d$),
$i,j$ denote the atomic sites and $\sigma$ the spin. 
$\epsilon^{\alpha\sigma}_{i}$ is the on-site energy which
can be written as 
$\epsilon^{\alpha}_{i}=\epsilon_0^{\alpha}+
\sigma h \delta_{\alpha {d}}$ with $h$ the
exchange energy and $\sigma=-1$ ($\sigma=+1$) for majority
(minority) spins. In equation (\ref{spham}),
$\gamma^{\alpha\beta\sigma}_{ij}=\gamma^{\alpha\beta}_{ij}$
is the hopping between the orbitals $\alpha$ and
$\beta$ at sites $i$ and $j$, and $c^\sigma_{\alpha i}$
($c_{\alpha i}^{\sigma\dagger}$) is the annihilation (creation)
operator for an electron at the atomic site $i$ in an orbital $\alpha$
with a spin $\sigma$. 
$h$ vanishes in the non-magnetic metal, and 
$\gamma^{\alpha\beta}_{ij}$ is zero if $i$ and $j$ do not
correspond to nearest neighbor sites.
Hybridization between the $s$ and $d$ orbitals is taken into
account by the non-vanishing term
$\gamma^{sd}$.
We have chosen to consider two orbitals per site in order to give an
appropriate description of the density of states of transition metals and
to take into account inter-band scattering
occurring at interfaces between different materials. The
DOS of a transition metal consists of a narrow 
band (mainly $d$-like) embedded in a broader band (mainly $sp$-like).
This feature can be reproduced in the above two band model, as shown
in reference \cite{noidd}, where the appropriate
choices for $\gamma^{\alpha\beta}_{ij}$
and $\epsilon^{\alpha}_{i}$
in Cu and Co are discussed.

We analyze the simplest generic model of disorder,
introduced
by Anderson within the framework of the localization theory \cite{and},
which consists of adding a random potential $V_i$
to each on-site energy, with a uniform distribution of width
$W$ ($-W/2\leq V \leq W/2$), centered on $V=0$

\begin{equation}
\tilde{\epsilon}^{\alpha\sigma}_{i}=\epsilon^{\alpha\sigma}_{i}+V
\;{.}
\label{and}
\end{equation}

\begin{figure}
\narrowtext
\epsfysize=8cm
\epsfxsize=7cm
\centerline{\epsffile{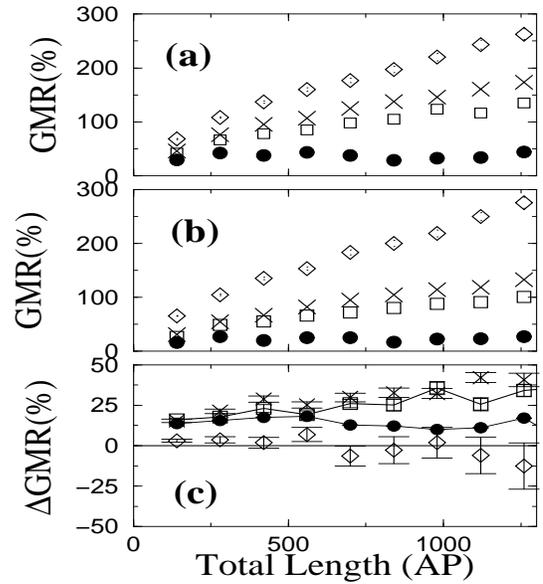}}
\caption{GMR for type I (a) and type II (b) multilayers,
and $\Delta$GMR (c) in the case of thin (10AP) 
and thick (40AP) Co layers, as a function of the number
of double bilayers Co/Cu/Co/Cu for different values
of disorder. The symbols represent respectively
$W=0$ ($\bullet$), $W=0.3$eV ($\Box$),
$W=0.6$eV ($\times$), $W=1.5$eV ($\Diamond$).}
\label{leed}
\end{figure}

The conductances and GMR ratios are calculated within the 
Landauer-B\"uttiker theory of transport \cite{but} using a technique already 
presented elsewhere \cite{noi}.
In figure \ref{leed} we present the mean GMR ratio for type I
(type II) multilayers GMR$_{\mathrm I}$ (GMR$_{\mathrm II}$) and the difference
between the GMR ratios of type I and type II multilayers 
$\Delta$GMR=GMR$_{\mathrm I}$-GMR$_{\mathrm II}$, 
as a function of $\mu$ for different values of the
on-site random potential. The average has been taken
over 10 different random configurations except
for very strong disorder where we have considered
60 random configurations. 
In the figure we display the standard deviation of the mean only
for $\Delta$GMR because for GMR$_{\mathrm I}$ and GMR$_{\mathrm II}$
it is negligible on the scale of the
symbols.
It is clear that type I 
multilayers possess a larger GMR ratio than type II multilayers, and
that both the GMR ratios and their difference increase
for large $\mu$.
These features are in agreement with experiments
\cite{chris2,msuord} and cannot be explained within the 
standard Boltzmann description
of transport.
The increase of the GMR ratio as a function of the number of 
bilayers is a consequence of enhancement of the
spin asymmetry of the current due to disorder. 
In fact, despite the Anderson potential being spin-independent
it will be more effective on the $d$ band
than on the $s$ band, because the former possesses a smaller
bandwidth.
Since the minority spin sub-band is dominated by the $d$-electrons
and the majority by the $s$-electrons, the disorder will suppress
the conductance more strongly in the minority band than in the majority.
Moreover, since the transport is phase-coherent, the asymmetry builds up
with the length, resulting in a length-dependent increase of the GMR ratio.
The different GMR ratios of type I and type II multilayers
can be understood by considering the inter-band 
scattering.
Both multilayers possess the same conductance in the P alignment,
while the conductance of type I multilayer in the AP alignment is
smaller than that of type II. The inter-band scattering is very
strong when an electron crosses phase-coherently a region
where the magnetizations have opposite orientations, and this
occurs in each (N/M/N/M$^\prime$) cell for type I multilayers,
while only in the central cell for type II multilayer 
(see figure \ref{ord-sc}a and \ref{ord-sc}b).
Hence the contribution to the resistance
in the AP alignment due to inter-band scattering is
larger in type I than in type II
multilayers.
Finally when the elastic mean free path is comparable with a single Co/Cu 
cell one expects the resistor model to become valid.
To illustrate this feature, figure \ref{leed} shows that in 
the case of very large disorder ($W=1.5$eV), 
$\Delta$GMR vanishes within a standard deviation
as predicted by the Valert and Fert theory.

\begin{figure}
\narrowtext
\epsfysize=5cm
\epsfxsize=7cm
\centerline{\epsffile{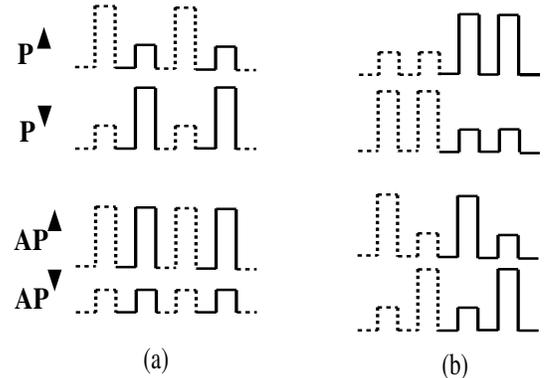}}
\caption{Heuristic scattering profiles for type I (a) and type II (b)
multilayers of the second example discussed in the text. The dashed and
continuous lines represent respectively scattering 
potentials of material M and M$^\prime$.
}
\label{prof}
\end{figure}

\begin{figure}
\narrowtext
\epsfysize=5cm
\epsfxsize=7cm
\centerline{\epsffile{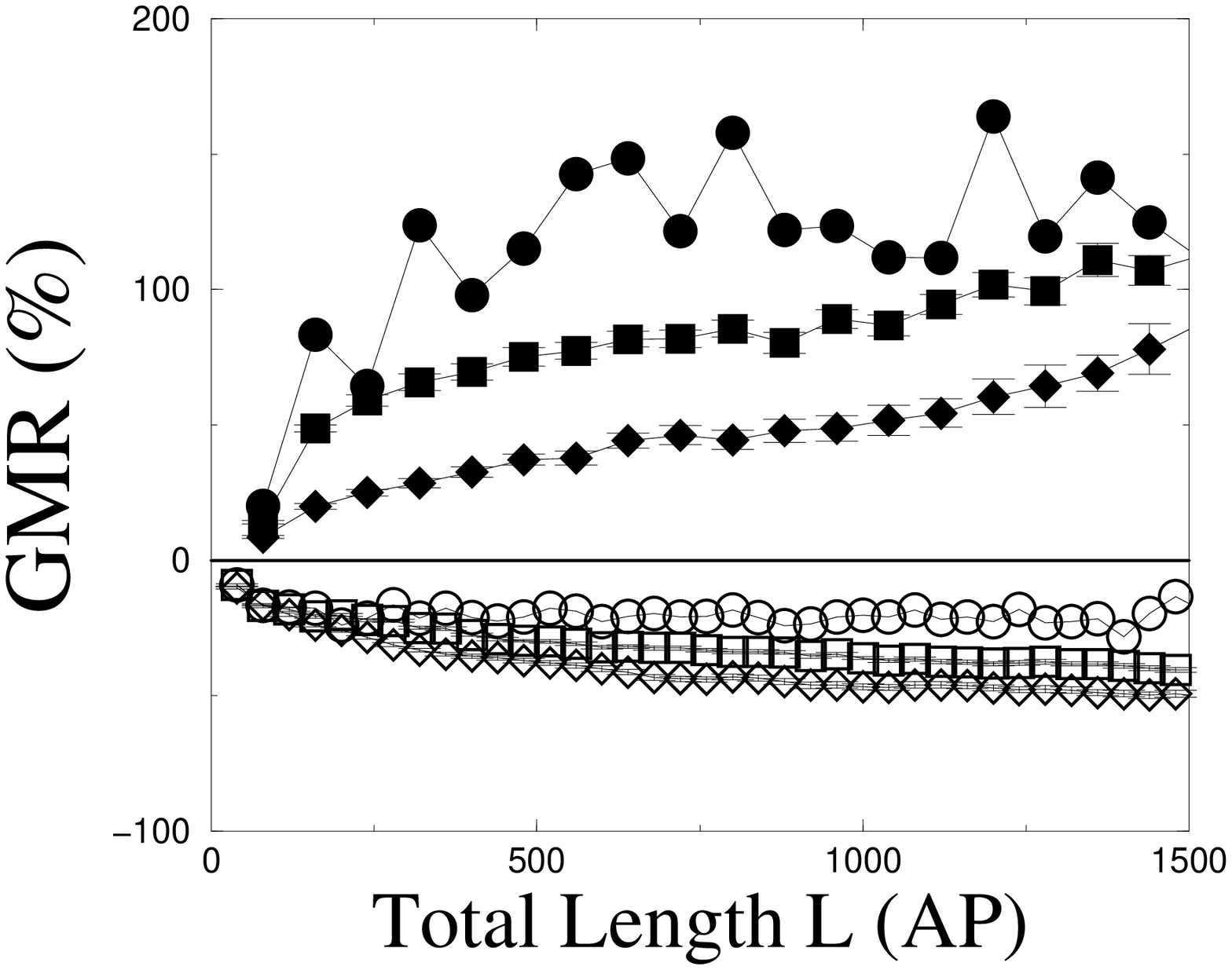}}
\caption{Different geometry-induced behavior of the GMR ratio as a function 
of disorder
in multilayers composed of Co and Fe$_{72}$V$_{28}$. In this case all the layer
thicknesses are fixed at 10AP.
The open (closed) symbols represent type I (type II) multilayers discussed
in the text. The circles are the disorder free case, squares and 
diamonds are for random on-site potentials of
0.6eV and 1.2eV respectively.}
\label{inverse}
\end{figure}

As a second example in which 
the dependence of the GMR ratio on disorder changes
when the multilayer geometry is varied, 
consider the system whose AP
alignment is sketched in figure \ref{ord-sc}c and \ref{ord-sc}d. 
In this case M and M$^\prime$ are different materials
chosen in such a way that the minority (majority) band of M possesses
a good alignment with the majority (minority) band of M$^\prime$.
Moreover the thickness of the N layers has been chosen in order to
allow an AP alignment of the magnetizations of adjacent magnetic
layers in both type I and type II multilayers. In this case both type I
and type II multilayers exhibit conventional P and AP alignments,
but their potential profile is quite different. In figure \ref{prof}
we present a schematic view of the potential profiles for type I and 
type II multilayers for both the spins in the P and AP configuration. 
A high barrier corresponds to large scattering and a small barrier
corresponds to weak scattering. The dashed line represents the effective
potential for material M and and the continuous line for material M$^\prime$.
Figure \ref{prof} illustrates that type I multilayers possess
a high transmission spin-channel in the AP alignment, and hence the 
resulting GMR ratio will be negative.
In contrast type II multilayers do not possess a high transmission
channel (there are large barriers for all spins in both
the P and AP configuration) and the sign of the GMR ratio will depend
on details of the band structure of M and M$^\prime$.
Consider the effects of disorder
on these two kinds of multilayers. Using
the same heuristic arguments as above we
expect that the GMR ratio of type I multilayers will increase
(become more negative) as disorder increases, in the case of disorder 
that changes the spin asymmetry of the current.
This is a consequence of the fact that,
in common with the conventional single-magnetic element,
one of the spin
sub-bands in the AP alignment is dominated by weak $s$-electrons
(small barrier), which are only weakly affected by disorder.
It is clear that this system is entirely
equivalent to conventional single-magnetic element
multilayers discussed above.
In contrast for type II multilayers there are no spin sub-bands
entirely dominated by the weak scattering (small barriers) $s$-electrons,
and all spins in either the P and AP configuration will undergo
scattering by the same number of high barriers.
In this case the effect of disorder will be
to increase all the resistances 
and this will result in a suppression of GMR.
Moreover it is important to note that in the 
completely diffusive regime, where the 
resistances of the different materials may be added in series, the GMR
ratio will vanish if $R_{\mathrm M}^{\uparrow(\downarrow)}\sim
R_{\mathrm M^\prime}^{\downarrow(\uparrow)}$, where
$R_{\mathrm A}^{\uparrow(\downarrow)}$ is the spin-dependent
resistance of the material A.
To verify this prediction we have simulated
both type I and type II multilayers using the parameters corresponding to
Co and Fe$_{72}$V$_{28}$ of reference \cite{noidd}, respectively 
for M and M$^\prime$, and
corresponding to Cu for N. 
This choice was motivated by the fact
that a reverse CPP-GMR has been obtained for 
(Fe$_{72}$V$_{28}$/Cu/Co/Cu)$\times\mu$ multilayers \cite{msufer}.
The GMR ratio for type I and type II multilayers
is shown in figure \ref{inverse}, which illustrates the remarkable result
that the GMR ratio of type I multilayers
increases with disorder, while for
type II structures it decreases. As explained above this is due to an enhanced asymmetry 
between the conductances in the P and AP alignment for type I
multilayers, and to a global increase of all the resistances for 
type II multilayers.
As far as we know there are no experimental studies of the
consequences of the geometry-dependent effect described above, 
and further investigation
will be of interest, in order to clarify the r\^ole of the disorder in 
magnetic multilayers.

Despite the fact that GMR was discovered more than ten years ago,
it continues to present fascinating insights into transport in
magnetic heterostructures. 
In this Letter we have addressed a new issue which lies outside the
widely-adopted Boltzmann description of GMR, namely that
changing the order of magnetic multilayers can
significantly alter the magnetoresistance \cite{chris2,msuord}.
We have shown that this effect is a consequence of phase coherence
on a length scale greater than the layer thicknesses.

\vspace{0.3in}

{\bf Acknowledgments}: The authors want to thank 
D.Bozec, C. Marrows, B.Hickey and M.Howson from the University of
Leeds for their suggestions and for the permission to discuss results not
yet published.
This work is supported by the EPSRC, the EU
TMR Programme and the DERA.

\end{multicols}

\end{document}